\documentclass[10pt,a4paper,twocolumn]{IEEEtran}
\normalsize
\usepackage[colorlinks,
linkcolor=black,
anchorcolor=black,
citecolor=black,
urlcolor=black,
]{hyperref}
\usepackage[hyphenbreaks]{breakurl}
\usepackage{diagbox}
\usepackage[english]{babel}
\usepackage{latexsym} 
\usepackage{varioref}
\usepackage{mathrsfs}
\usepackage{array}
\usepackage{hhline} 
\usepackage{dcolumn}
\usepackage{tabularx}
\usepackage{algorithm}
\usepackage{algorithmic}
\usepackage{calc}
\usepackage{epsfig}
\usepackage[usenames]{color}
\usepackage{amsmath,amssymb,bigdelim}
\usepackage{t1enc}
\usepackage{pstricks,pst-node,psfrag,pst-plot}
\usepackage{pstool} 
\usepackage{graphicx,parskip,amsbsy}
\usepackage{boxedminipage}
\usepackage{indentfirst}
\usepackage{booktabs}
\usepackage[noadjust]{cite}
\usepackage{caption}
\usepackage{multirow}
\usepackage{subfigure}
\usepackage[usenames]{color}
\usepackage{clrscode}
\usepackage{url}
\usepackage{setspace}
\usepackage{fancybox}
\usepackage{ulem}
\makeatletter
\def\url@leostyle{%
  \@ifundefined{selectfont}{\def\UrlFont{\sf}}{\def\UrlFont{\small\ttfamily}}}
\makeatother
\newcommand{\rem}[1]{}

\newcommand{\addnew}[1]{{\color{red} #1}}
\newcommand{\rmold}[1]{}

\renewcommand{\baselinestretch}{1}
\providecommand{\texlivekeywords}[1]{\textbf{\textit{Index terms---}} \textbf{\small #1}}
\def\BibTeX{{\rm B\kern-.05em{\sc i\kern-.025em b}\kern-.08em
    T\kern-.1667em\lower.7ex\hbox{E}\kern-.125emX}}

\begin{document}

	\pagestyle{plain}
  \title{A Centralized and Scalable Uplink Power Control Algorithm in Low SINR Scenarios}

	\author{Xuesong Cai, Istv\'an Z. Kov\'acs, Jeroen Wigard, and Preben E. Mogensen

		\thanks{
    This work was supported by Nokia. \textit{(Corresponding author: Xuesong Cai).}

    X. Cai is with the Department of Electronic Systems, Aalborg University,
    Aalborg 9220, Denmark and with the Department of Electrical and
    Information Technology, Lund University, 22100 Lund, Sweden (email:
    xuesong.cai@ieee.org).

I. Z. Kov\'acs and J. Wigard are with Nokia Standards, 9220 Aalborg, Denmark (email: istvan.kovacs@nokia.com; jeroen.wigard@nokia.com).

P. E. Mogensen is with the 
Department of Electronic Systems, Aalborg University, 9220 Aalborg, Denmark, and Nokia Standards, 9220 Aalborg, Denmark (e-mail:
pm@es.aau.dk).

}

	}

\markboth{IEEE Transactions on Communications}%
{Submitted paper}

\maketitle \thispagestyle{plain}

  

  

\begin{abstract}

  \black{Power control is becoming increasingly essential for the fifth-generation (5G) and beyond systems. An example use-case, among others, is the unmanned-aerial-vehicle (UAV) communications where the nearly line-of-sight (LoS) radio channels may result in very low signal-to-interference-plus-noise ratios (SINRs).}
  \rmold{With the introduction of aerial user equipments in cellular networks, resulting in an increase of line of sight links, power control is becoming more and more vital to enable 
   the (uplink) high-throughput data streaming and protect the users on the ground.} {Investigations in \cite{gppower} proposed to efficiently and reliably solve this kind of {non-convex problem} via a series of geometrical programmings (GPs) using condensation approximation.} {However, it is only applicable for a small-scale network with several communication pairs and practically infeasible with more (e.g. tens of) nodes to be jointly optimized.} 
   {We therefore in this paper aim to provide new insights into this problem. By properly introducing auxiliary variables, the problem is transformed to an equivalent form which is simpler and more intuitive for condensation. A novel condensation method with linear complexity is also proposed based on the form. 
   The enhancements make the GP-based power control feasible for both small- and especially large-scale networks that are common in 5G and beyond. The algorithm is verified via simulations. A preliminary case study of uplink UAV communications also shows the potential of the algorithm.}

\end{abstract}
\texlivekeywords{Interference management, power control, geometrical programming, uplink and UAV.}
\IEEEpeerreviewmaketitle
\section{Introduction}
Interference management has been investigated for decades. Many works, e.g. \cite{MAPEL,RR,gppower}, have shown that significant gains can be achieved through power control. As the network becomes densified and different types of user equipments are being involved in the fifth generation (5G) communication networks and beyond, interference has been considered a major limiting factor of the system. For example, in the unmanned-aerial-vehicle (UAV) communications, interferences of both up- and down-link become severe with height increasing, due to the close-to line-of-sight (LoS) links between UAVs and terrestrial base stations (BSs). 
The power control problem usually has the form of maximizing the weighted-sum-rate of the system, with each receiver node (Rx) satisfying its power and quality of service (QoS) constraints. This is generally a non-convex problem and difficult to obtain the global optimality. Different algorithms have been proposed. Without considering the QoS constraints, the ADP (Asynchronous Distributed Pricing) algorithm was proposed in \cite{ADP} where each Rx sends out a price and updates its transmitting power according to the prices sent by other links iteratively until convergence. In \cite{RR}, the power allocation was obtained by tuning the current link's power while fixing other links' transmission power to maximize the system capacity in a Round-Robin (RR) manner (one by one) until convergence. In \cite{BPC}, the authors proposed to utilize binary power control (i.e. either transmitting with zero power or maximum power), and results show that the performance loss to global optimality is insignificant. When considering QoS constraints the problem becomes more difficult.\footnote{The problem without QoS constraints can be considered a special case with QoS constraints.} 
In \cite{MAPEL}, by iteratively shrinking the polyblock, the proposed MAPEL algorithm can asymptotically approach the global optimality, although its complexity increases significantly with the number of link pairs increasing. In \cite{DNN}, the authors exploited the recent advances in deep learning and proposed an ensemble deep-neural-network to tackle the problem. {In \cite{gppower}, the authors approximated the problem as a geometrical programming (GP) in the high signal-to-interference-plus-noise (SINR) regime. This method is arguably the best algorithm since GP can be solved efficiently and reliably \cite{cvx}. {Thus, we focus on GP-based power control in this paper.} \black In addition, a more comprehensive review of different power control algorithms can also be found in \cite{DNN} and references therein.} 

However, in the low SINR regime, the convex-approximation in \cite{gppower} is invalid. Therefore, a condensation method was also proposed in \cite{gppower} to solve the original problem via a series of GP problems. Nevertheless, the condensation is performed at  power variables, which is non-straightforward and non-scalable.\rmold{In other words, it is practically infeasible/impossible to use the condensation method proposed in \cite{gppower} for a larger-scale network even with not so many link pairs, which will be further discussed in Sect.\,\ref{sect:cond}.} {In other words, it is even practically infeasible for a relatively large network (which will be further discussed in Sect.\,\ref{sect:cond}), although optimizing a moderate to large network is inevitable for 5G with network densification and different types of user equipments involved, e.g., UAVs.} {This is where this paper will provide new insights and enhancements. By introducing auxiliary variables, the problem is more intuitively interpreted. A novel condensation approximation method is also proposed by leveraging the auxiliary variables, so that the number of parameters to be calculated increases linearly along the number of links. The enhancements make the GP-based algorithm applicable for both small and especially large-scale networks that are common in 5G and beyond communications. Moreover, a preliminary case study for uplink UAV communications is also conducted to verify the algorithm as well as illustrate the potential of the algorithm when applied in 5G and beyond communications.}
\rmold{
To solve the problem, the contributions of this paper 
are mainly three-folds. \textit{1)} A standard form of the original problem is proposed by introducing auxiliary variables. In the standard form, the condensation can be applied for auxiliary variables which is more intuitive. \textit{2)}  A new condensation method is proposed, where the number of parameters to be calculated increases linearly with the number of links. Moreover, the proposed method is more straightforward as there is no coupling among those auxiliary variables compared to directly conducting the condensation for the power variables. The method can be easily scaled for large-scale networks. \textit{3)} In addition, by using the proposed method, a case study for up-link UAV communications in a (moderately) large-scale cellular network is also illustrated.
}The rest of the paper is organized as follows. Sect.\,\ref{section:problem} elaborates the problem formulation and transformation via introducing auxiliary variables. Sect.\,\ref{sect:cond} discusses the condensation principle and proposes the novel condensation method. The {algorithm verification} and case study are presented in Sect.\,\ref{sect:case}. Finally, conclusive remarks are included in Sect.\,\ref{sect:conclusions}

\section{Problem formulation\label{section:problem}}
Let us consider the power control problem in a wireless network with a set of $\mathcal{N}=\{1,\cdots,N\}$ of distinct link pairs (e.g., Fig.\,\ref{fig:cells}). Each link pair has a transmitter node (Tx) and a Rx. The channel gain matrix is denoted as $\mathbf{G}=[G_{ij}]$ with $G_{ij}$ indicating the channel gain between the $i$th Tx and the $j$th Rx. Note that $G_{ij}$ is attributed to path loss, shadowing, fast fading, etc. The node pair $(i,i)$ and node pairs $(i,j), j\neq i$ are the serving link and interfering links, respectively. The transmit power $p_i$ at the $i$th Tx is usually bounded between $p_{i,\text{min}}$ and $p_{i,\text{max}}$. Moreover, the noise power measured at the
$i$th Rx is denoted as $n_i$. Therefore, the received SINR $\gamma_i$ at the $i$th Rx can be calculated as
\begin{equation}
\begin{aligned}
\gamma_i(\mathbf{p}) = \frac{p_i G_{ii}}{{n_i} + \sum_{j\in \mathcal{N}, j\neq i} p_j G_{ji} }
\end{aligned}.
\end{equation} where $\mathbf{p}=[p_1,\cdots, p_N]$ is the compact vector notation of the transmitted power of all the Txs. We consider the data rate $R_i$ (bit/sec/Hz) at the $i$th Rx node according to the modified Shannon capacity formula as
\begin{equation}
\begin{aligned}
R_i(\mathbf{p}) = a\log_{2}(1+b\gamma_i)
\end{aligned}.
\label{eq:rate_calculation}
\end{equation} where $a$ and $b$ are constants no greater than 1. This is caused by different factors such as the coding gap to Shannon capacity, system efficiency, etc., and has been certified in \cite{preben_lte_capacity} in LTE networks. Note that With $a$ and $b$ as 1, \eqref{eq:rate_calculation} becomes the Shannon capacity formula.

The objective of power control is to find the optimal transmitted power $\mathbf{p}^{\ast}$ that leads to the maximum weighted sum rate for the whole network with possible QoS constraints for individual link pairs. The optimization problem can be formulated as
\begin{equation}
\begin{aligned}
\text{maximize}\quad & \sum_{i \in \mathcal{N}} w_i R_i \quad\quad {\text{w.r.t.} \quad \mathbf p}\\
\text{subject to} \quad & R_i \geq R_{i,\text{min}}, \forall i \in \mathcal{N} \\
  & p_{i,\text{min}} \leq p_i \leq  p_{i,\text{max}}, \forall i \in \mathcal{N}
\end{aligned}
\label{eq:original_problem}
\end{equation} where $w_i$ is the weight (importance) for the $i$th Rx node, and $R_{i,\text{min}}$ is the QoS constraints for the $i$th Rx (which can be formulated equivalently as $\gamma_i \geq \gamma_{i,\text{min}}$).  As a special case with $\mathbf{R}_{\text{min}} = [R_{i,\text{min}}, \cdots, R_{N,\text{min}}]$ as $\textbf{0}$, the maximization problem
\eqref{eq:original_problem} becomes an unconstrained problem in terms of QoS. By introducing auxiliary variables $\mathbf{s}=[s_1,\cdots,s_N]$ and $r$, we can further equivalently transform \eqref{eq:original_problem} to
\begin{equation}
\begin{aligned}
\text{maximize}\quad &  r   \quad\quad {\text{w.r.t.} \quad \mathbf p, \mathbf s}\\
\text{subject to} \quad & \prod _{i=1}^{N} (1+s_i)^{w_i} \geq r  \\
& \frac{p_i G_{ii}}{{n_i} + \sum_{j\in \mathcal{N}, j\neq i} p_j G_{ji}} \geq s_i , &\forall i  \\
& s_i \geq \gamma_{i,\text{min}} , &\forall i  \\
  & p_{i,\text{min}} \leq p_i \leq  p_{i,\text{max}}, &\forall i
\end{aligned}.
\label{eq:transformed_problem1}
\end{equation}
One step further, we have
\begin{equation}
\begin{aligned}
\text{minimize}\quad &  r^{-1}   \quad\quad {\text{w.r.t.} \quad \mathbf p, \mathbf s}\\
\text{subject to} \quad & \frac{r}{\prod _{i=1}^{N} (1+s_i)^{w_i}} \leq 1  \\
& s_i p_i^{-1} G_{ii}^{-1} ({n_i} + \sum_{j\in \mathcal{N}, j\neq i} p_j G_{ji}) \leq 1, &\forall i  \\
& s_i^{-1} \gamma_{i,\text{min}} \leq 1, &\forall i  \\
  & p_i^{-1}  p_{i,\text{min}} \leq 1, &\forall i   \\
    &  p_i p_{i,\text{max}}^{-1}\leq 1 , &\forall i.
\end{aligned}
\label{eq:transformed_problem2}
\end{equation} With the above transformation introducing nonnegative auxiliary variables $\mathbf s$ and $r$, we consider \eqref{eq:transformed_problem2} a standard form of the weighted sum rate maximization problem. The optimal power allocation $\mathbf{p}^{\ast}$ is obtained when achieving the minimum $r^{-1}$ (i.e., $r^{\ast}$), and the maximum weighted sum rate can be calculated as $\log_{2}r^{\ast}$. {It is worth noting that the {transformed} form
\eqref{eq:transformed_problem2} is essential for the proposed condensation approximation in Sect.\,\ref{sect:cond}, since it provides an alternative and intuitive way by leveraging the auxiliary variables $\mathbf s$.}
\rmold{It is worth noting that the \addnew{transformed} form
\eqref{eq:transformed_problem2} is essential for the condensation in a large-scale network in the low SINR regime, as we can only consider the condensation for $\mathbf s$ rather than $\mathbf p$ which will be discussed in Sect.\,\ref{sect:cond}.}

\section{Condensation method in the low SINR regime\label{sect:cond}}
Before going to the low SINR regime, let us first consider the high SINR regime. In the high SINR regime, the maximum weighted sum rate is considered to be achieved with all the Rxs have high SINRs, which means that $s_i^{\ast}$ or $\gamma_i^{\ast}, \forall i$ is (much) larger than 1. Therefore, the term $\frac{1}{\prod _{i=1}^{N} (1+s_i)^{w_i}}$ in \eqref{eq:transformed_problem2} can be well approximated as $\prod_{i=1}^{N} s_i^{-w_i}$ so that
\eqref{eq:transformed_problem2} becomes a standard GP problem where a posynomial is to be minimized subject to upper bounded posynomial constraints and equality monomial constraints \cite{gp,gppower}. Briefly, a monomial has the form as
\begin{equation}
\begin{aligned}
f(\mathbf{x}) = c x_1^{d_1} x_2^{d_2}\dots x_n^{d_n}
\end{aligned}.
\end{equation} where $x_i$'s and $c$ are nonnegative variables and constant, respectively, and $d$'s are real constants. A posynomial has the form as the sum of several monomials. In the high SINR regime, the GP can be efficiently and numerically reliably solved using, e.g. the interior point method \cite{cvx}, to find the global optimal $\mathbf{p}^\ast$.

However, in the low SINR regime with severe interference, the approximation as done in the high SINR regime is not valid anymore, and obviously $\frac{1}{\prod _{i=1}^{N} (1+s_i)^{w_i}}$ is not a posynomial. Therefore, a condensation method was proposed in \cite{gppower} to solve a series of GP problems to find the power allocation satisfying the Karush–Kuhn–Tucker (KTT) conditions (which means that the final power allocation could be a local maxima) in the low SINR regime. The basic idea is to approximate the non-posynomial term in the denominator as a monomial at a given feasible power allocation point, get a new optimal power allocation for the currently approximated GP, then approximate the original problem at the new power allocation again to further get another power allocation. The process is proceeded iteratively until convergence. The monomial approximation proposed in \cite{gppower} is based on the arithmetic-geometric mean inequality. Specifically, the approximated monomial $\hat{g}(\mathbf{x})$ for a function $g(\mathbf{x})=\sum_i u_i(\mathbf{x})$ can be written as
\begin{equation}
\begin{aligned}
\hat{g}(\mathbf{x}) = \prod_i (\frac{u_i(\mathbf x)}{\alpha_i})^{\alpha_i}
\end{aligned}
\label{eq:old_condensation}
\end{equation}
where $u_i$ is a monomial component, and $\alpha_i$ is calculated as  $\frac{u_i(\mathbf{\mathbf x_0})}{g(\mathbf x_0)}$ at the approximation point $\mathbf x_0$. Furthermore, $\hat{g}(\mathbf x)$ has to satisfy three conditions \cite{gppower,MarksInner} to guarantee the power allocation converge to a KTT point\footnote{{The convergence point may be different depending on the power initialization.}}, which include: \textit{(a)} $\hat{g}(\mathbf x) \leq g(\mathbf x)$ for all $\mathbf x$. This is to tighten the constraint so that the obtained new power allocation for the current approximated GP is always feasible for the original problem. \textit{(b)} $\hat{g}(\mathbf x_0) = g(\mathbf x_0)$.
This is to guarantee the monotonicity of the optimal values obtained in successive iterations. \textit{(c)} $\nabla \hat{g}(\mathbf x_0) = \nabla g(\mathbf x_0)$.
This is to guarantee the KTT conditions for the original problem are satisfied after convergence. The condensation \eqref{eq:old_condensation} proposed in \cite{gppower} satisfies the three conditions as discussed in \cite{gppower}, and simulations have shown its performance, e.g. in a small-scale network with 3 link pairs in \cite{gppower} and up to 10 link pairs in \cite{MAPEL}. Nevertheless, we would like to note that there is a major problem when \eqref{eq:old_condensation} is applied in a large-scale network with  a certain number of link pairs. The reason is that to conduct condensation
\eqref{eq:old_condensation}, one has to firstly rewrite $g(\mathbf x)$ in the form of summing several monomials. For a small network, this could be done practically. However, the number of monomial terms increasing exponentially with the link number increasing, which means that it is practically difficult to conduct \eqref{eq:old_condensation} in a larger scale network. As an example, considering the term  $g(\mathbf s) = {\prod _{i=1}^{N} (1+s_i)^{w_i}}$, it has $2^{N}$ monomial terms. With 20 link pairs, there will be more than one million monomial terms meaning more than one million $\alpha$'s have to be calculated. Moreover,
$2^N$ is for the variables $\mathbf s$ in the form \eqref{eq:transformed_problem2} as proposed in this work. With the condensation applied directly for power variables ($g(\mathbf p) = \prod_{i=1}^{N} (1+\frac{p_i G_{ii}}{{n_i} + \sum_{j\in \mathcal{N}, j\neq i} p_j G_{ji}})^{w_i}$
as done in \cite{gppower}), the number of monomial $u_i$'s increases much faster than $2^N$. Thus, a scalable condensation method that can be applied for a larger-scale network is in necessity for 5G and beyond communications.

As the proposed form \eqref{eq:transformed_problem2} is general, we only need to focus on the condensation for auxiliary variables
\begin{equation}
\begin{aligned}
g(\mathbf s)= {\prod_{i=1}^{N} (1+s_i)^{w_i}}
\end{aligned}
\label{eq:gs}
\end{equation}
Before proposing the final condensation for $g(\mathbf s)$ , we firstly see the function $h_i(s_i) = 1+s_i$. Considering {that we want to find} a monomial $\hat{h}_i (s_i)= c_i s_i^{d_i}$ that satisfies the conditions \textit{(b)} and \textit{(c)} with $h_i$ at a given $s_{i0}$, we have
\begin{equation}
\begin{aligned}
  & \hat{h}^\prime_i(s_i) \Biggr|_{s_{i0}}=  h_i^{\prime}(s_i)\Biggr|_{s_{i0}}, \hat{h}_i(s_{i0}) = h_i(h_{i0}).
\end{aligned}
\end{equation}
which is equivalently as
\begin{equation}
\begin{aligned}
\frac{d}{{d}s_i} \ln(\hat{h}_i) \Biggr|_{s_{i0}}= \frac{d}{{d}s_i} \ln(h_i)\Biggr|_{s_{i0}}, \hat{h}_i(s_{i0}) = h_i(s_{i0}).
\end{aligned}
\label{eq:aa}
\end{equation}
According to \eqref{eq:aa}, it is straightforward to find $d_i = \frac{s_{i0}}{1+s_{i0}}$ and $c_i = (1+s_{i0})  s_{i0}^{-d_i}$. To show that $h_i$ and $\hat{h}_i$ satisfy the condition \textit{(a)}, i.e. $\hat{h}_i(s_i) \leq h_i(s_i)$ for all $s_i$, we construct the difference function
\begin{equation}
\begin{aligned}
  l(s_i) = \hat{h}_i(s_i) - h_i(s_i) \\
\end{aligned}
\end{equation} It can be calculated that
\begin{equation}
\begin{aligned}
  &  l^\prime(s_i) = (\frac{s_{i0}}{s_i})^{\frac{1}{1+s_{i0}}} - 1, & \Downarrow  \\
  & \ln( l^\prime (s_i)) = \frac{1}{1+s_{i0}} \ln(\frac{s_{i0}}{s_i})
\end{aligned}
\end{equation} where when $s > s_{i0}$, $l^\prime$ is negative, and vice versa. Thus, $l(s_i)$ is maximized at $s_{i0}$ as 0, which means that condition \textit{(a)} holds for $\hat{h}_i(s_i)$ and $ h_i(s_i)$. Finally we can write the condensation function $\hat{g}(\mathbf s)$ for $ g(\mathbf s)$ at $\mathbf s_0 = [s_{10}, \dots, s_{N0}]$ as
\begin{equation}
\begin{aligned}
\hat g (\mathbf s)= {\prod_{i=1}^{N} c_i^{w_i} s_i^{w_i d_i}}.
\end{aligned}\label{eq:13}
\end{equation} It can be known that $\hat{g}(\mathbf s)$ and $g(\mathbf s)$ satisfy conditions \textit{(a)} and \textit{(b)}, since each $\hat{h}_i(s_i)$ and $  h_i(s_i)$ satisfy conditions \textit{(a)} and \textit{(b)}. Condition \textit{(c)} also holds for $\hat{g}(\mathbf s)$ and $g(\mathbf s)$, which can be directly checked by comparing their gradients. It is worth noting that the calculation for a $d_i$ is only related to $s_i$ as
$d_i=\frac{s_{i0}}{1+s_{i0}}, \forall i$ (decoupled from all the other $s_j ,j \neq i$), and the multiplicative constant
$c = \prod_{i=1}^{N} c_i^{w_i}$ can be calculated directly as $c = g(\mathbf s_0)(\prod_{i=1}^{N} s_{i0}^{w_i d_i})^{-1}$ after all $d_i$'s are obtained. This means that the proposed condensation method is easy and straightforward to be done.

To conclude, by exploiting the standard form as transformed in
\eqref{eq:transformed_problem2}, the condensation method in \eqref{eq:13} is proposed for the general power control problem. Furthermore, the number of calculated parameters in the condensation scales linearly with $N$ (which is actually $N+1$). This method makes the power control problem in the low SINR regime be practically solvable using a series of GPs for both small-scale and (very) large-scale networks. The pseudocode in Algorithm\,\ref{al1} illustrates the process for the problem \eqref{eq:transformed_problem2} using the novel condensation method.\footnote{An initial feasible $\mathbf p$ with QoS constraints can be found using the method as discussed in Sect.\,III-B in \cite{DNN}. Without QoS constraints, setting an initial feasible $\mathbf p$ is trivial. {Moreover, simulations have shown the algorithm can converge within several iterations even for very large networks, e.g. the one as illustrated in Fig.\,\ref{fig:cells} and {Fig.\,3(c)}. With the proposed condensation approximation of linear complexity, Algorithm\,\ref{al1} keeps the polynomial time complexity of the interior-method. Whereas the exponential complexity of the previous condensation method dominants the algorithm complexity when the network scales to a large one.}}

\begin{algorithm}
\setstretch{1}
{\textbf {Input}: An initial feasible power allocation $\mathbf p$.\\} 
{\textbf {Output}: A power allocation that satisfies KKT conditions for problem \eqref{eq:transformed_problem2}.}

\begin{algorithmic}[1]
\STATE \textbf{Repeat:}
\STATE Calculate $\mathbf s_0$ as $\mathbf s_0 = [\gamma_i(\mathbf p),\dots, \gamma_N(\mathbf p)]$.
\STATE Conduct the condensation proposed in \eqref{eq:13} at $\mathbf s_0$ for $g(\mathbf s)$ in \eqref{eq:gs}.
\STATE Solve the resulted GP problem using the interior method, and update $\mathbf p$ as $\mathbf p^{\ast}$ obtained in this step.
\STATE \textbf{Until} the power difference between two successive iterations satisfies $||\mathbf p_{\text{new}} - \mathbf p_{\text{old}} || < \epsilon$ with $\epsilon$ a pre-defined tolerance.
 \caption{Solving the power control problem \eqref{eq:transformed_problem2} using the proposed condensation method \eqref{eq:13}. \label{al1}}
\end{algorithmic}
\end{algorithm}

\section{performance evaluation and Cases study \label{sect:case}}
\subsection{Case 1: Probability of achieving global optimality} 
\begin{figure}
\begin{center}
  
\psfrag{Simulation Runs}[c][c][0.6]{Simulation runs index}
\psfrag{Obtained wighted sum rate}[c][c][0.6]{Weighted-sum-rate [bit/s/Hz]}
\psfrag{data1}[l][l][0.6]{Algorithm\,\ref{al1}}
\psfrag{algorithm111}[l][l][0.6]{Algorithm\,\ref{al1}}
\psfrag{data2}[l][l][0.6]{MAPEL}
\includegraphics[width=0.45\textwidth]{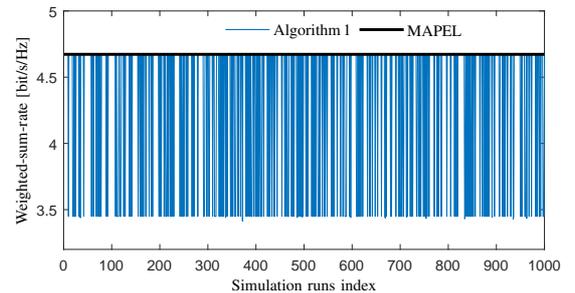}
    
\end{center}
\caption{The weight sum rate obtained for the case of \textbf{Example 1} in \cite{MAPEL} using Algorithm\,\ref{al1} and MAPEL, respectively. \label{fig:probability}}
\end{figure}

In the low SINR regime, Algorithm\,\ref{al1} not necessarily converges to the global optimal power allocation. To study the probability, the ground truth of global optimum has to be obtained. Here, we resort to the MAPEL algorithm in \cite{MAPEL}. Note that although MAPEL can obtain the globally optimal power allocation, its computation complexity increases drastically with the network size increasing \cite{MAPEL,DNN}. Thus we choose the same small-scale network as the \textbf{Example 1} presented in \cite{MAPEL}, which is a network with four link pairs. The channel gain matrix is
\begin{equation}
G={
\left[ \begin{array}{cccc}
0.4310 &   0.0002 &  0.2605 &  0.0039\\
  0.0002 & 0.3018 &  0.0008 &  0.0054\\
  0.0129 &  0.0005 &  0.4266 &  0.1007\\
  0.0011 &  0.0031 &  0.0099 &  0.0634
\end{array}
\right ]},
\end{equation} power upper bounds are $[0.7, 0.8, 0.9, 1.0]$\,mW, noise power is 0.1\,$\mu$W for all links, and the weights are $[\frac{1}{6}, \frac{1}{6}, \frac{1}{3}, \frac{1}{3}]$. Fig.\,\ref{fig:probability} illustrates the obtained weighted sum rate using Algorithm\,\ref{al1} with 1000 random power initializations, and the black horizontal line indicates the global maximum weighted sum rate obtained by using the MAPEL algorithm.
The probability of Algorithm\,\ref{al1} achieving the global optimality is calculated as 73.4\% in this case, which is slightly larger than 70.8\% presented in \cite{MAPEL} when using the condensation method \eqref{eq:old_condensation}.

\subsection{Case 2: Close-to real-world up-link (UL) UAV communications in cellular networks}
\begin{figure}
\begin{center}
\psfrag{BSs}[l][l][0.6]{BSs}
\psfrag{UAVs}[l][l][0.6]{UAVs}
\includegraphics[width=0.3\textwidth]{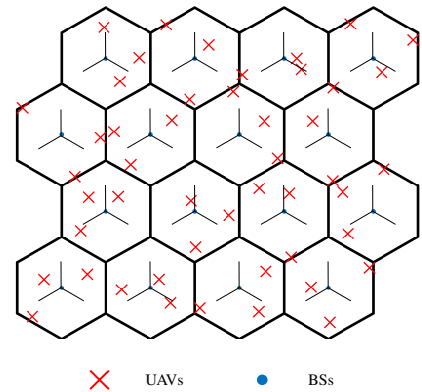}
\end{center}
\caption{An example  {realization of 48 UAVs located} in the sectorized cells in the case study for UL transmission of cellular-UAVs.\label{fig:cells}}
\end{figure}

Recently, UAV is gaining its popularity in multiple applications due to its low cost and flexibility \cite{7936620,8576578}. The cellular networks, e.g. LTE, are considered promising to provide critical and non-critical communications to UAVs. Nevertheless, due to the clearance of the channel between UAVs and terrestrial BSs \cite{7936620,8576578,cai2021empirical}, both the down-link and UL experience severe interference \cite{8807190,8369158,8287891}, which limits the system capacity significantly. To preliminarily gain insights into the UL communication of cellular-UAVs {and show the potential of the proposed algorithm}, we study the power control for UAVs in a cellular network.\footnote{{Some other interesting use-cases of optimization can also be found in \cite{8415781,9115248,9121255}. Joint power control and beamforming problem was addressed for a typical two-user uplink mm-wave-NOMA system in \cite{8415781}. In \cite{9115248}, a mm-wave system with UAV as a relay was considered and optimized via proper beamforming, power control and positioning of the UAV. In \cite{9121255}, the placement of UAV-BSs and resource allocation aiming energy-efficient internet-of-things was investigated.}} As illustrated in Fig.\,\ref{fig:cells}, a network with 48 cells ($4\times4$ sectorized hexagons) is considered in the simulation.\footnote{The number of monomial terms in \eqref{eq:old_condensation} is much higher than $2^{48}$.} The distance between neighboring BSs is set as 2\,km, and the heights of BSs are 35\,m. In each sector the half power beam-widths (HPBWs) of the sector antenna in azimuth and elevation domains are set as
120$^\circ$ and 13$^\circ$, respectively, and the down-tilt angle is properly set (as 8.5$^\circ$ in this case) to optimize the ground coverage. An UAV with 60\,m height is randomly put in each cell. The maximum transmission power of an UAV is set as 23 dBm, the noise power spectrum density is calculated at 290\,K, and the weights are set identical for all UAVs as
$\omega_i = \frac{1}{N}, \forall i$ meaning that the weighted sum rate is the average value for all UAVs. In addition, we assume the UAVs are using omnidirectional antennas. The channel model is from the results in \cite{7936620}.\footnote{Fast fading is not considered in this case, as the channel with UAV in the sky has a large K-factor \cite{Xoselte}. On the other hand, we are not attempting to accurately reproduce the channel, and the obtained results can be considered as an upper-bound performance.}  Table\,\ref{tab:parameters} summarizes the important parameters configured in the case study.

\begin{table}[t]
\centering
\caption{The simulation configuration of the case study for UL UAV communications.}
\scalebox{1}{
\begin{tabular}{cc}
\hline
\hline
\multicolumn{2}{c}{{{{\textit{Main parameters in the simulation}}}}}\\
\hline
Network scale  & 48\,cells      \\
Cell type & Sectorized hexagon \\
BS spacing & 2\,km \\
BS height & 35\,m \\
HPBWs of sector antenna & (120$^\circ$, 13$^\circ$) \\
Down-tilt angle & 8.5$^\circ$ \\
UAV height & 60\,m \\
Max. transmit power per UAV-UE & 23\,dbm\\
{\black Schedule assumption }  & \begin{tabular}[c]{@{}c@{}}\black One UAV-UE/cell/TTI \\\end{tabular}\\
\hline\hline
\end{tabular}}
\label{tab:parameters}
\end{table}
 
\begin{figure}
\begin{center}
\psfrag{Run index}[c][c][0.6]{Realization index}
\psfrag{rate}[c][c][0.6]{Average rate [bit/s/Hz]}
\psfrag{data2}[l][l][0.6]{Algorithm\,\ref{al1}, no QoS constraint}
\psfrag{data3}[l][l][0.6]{Algorithm\,\ref{al1}, OLPC QoS constraints}
\psfrag{data1}[l][l][0.6]{3GPP OLPC}
\subfigure[]{\includegraphics[width=0.42\textwidth]{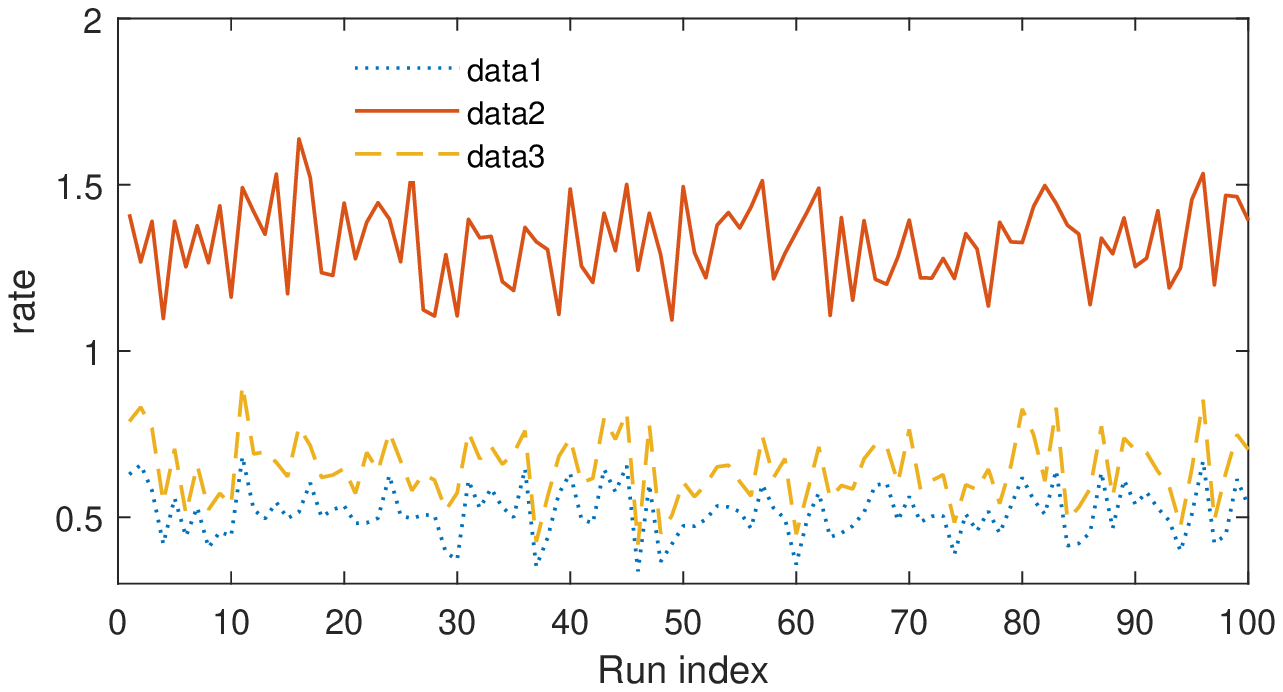}\label{fig:avepfm}}
        \psfrag{data2}[l][l][0.6]{Algorithm\,\ref{al1}, no QoS constraint}
\psfrag{data3}[l][l][0.6]{Algorithm\,\ref{al1}, OLPC QoS constraints}
\psfrag{data1}[l][l][0.6]{3GPP OLPC}
\psfrag{x}[c][c][0.6]{Rate [bit/s/Hz]}
\psfrag{F(x)}[c][c][0.6]{$\mathrm{P}(\text{Rate} \leq abscissa)$}
\psfrag{z}[c][c][0.2]{Rate [bit/s/Hz]}
\psfrag{y}[c][c][0.2]{$\mathrm{P}(\text{Rate} \leq abscissa)$}
\subfigure[]{\includegraphics[width=0.42\textwidth]{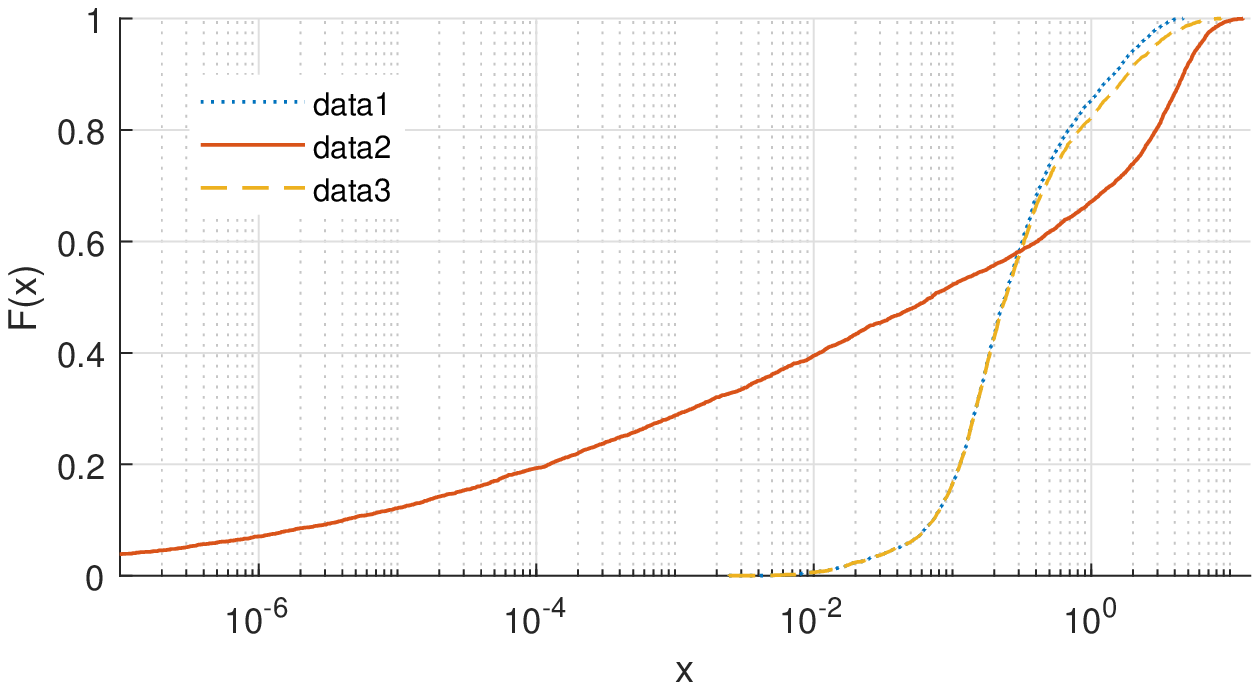}\label{fig:cdfs}}
    \subfigure[]{\includegraphics[width=0.42\textwidth]{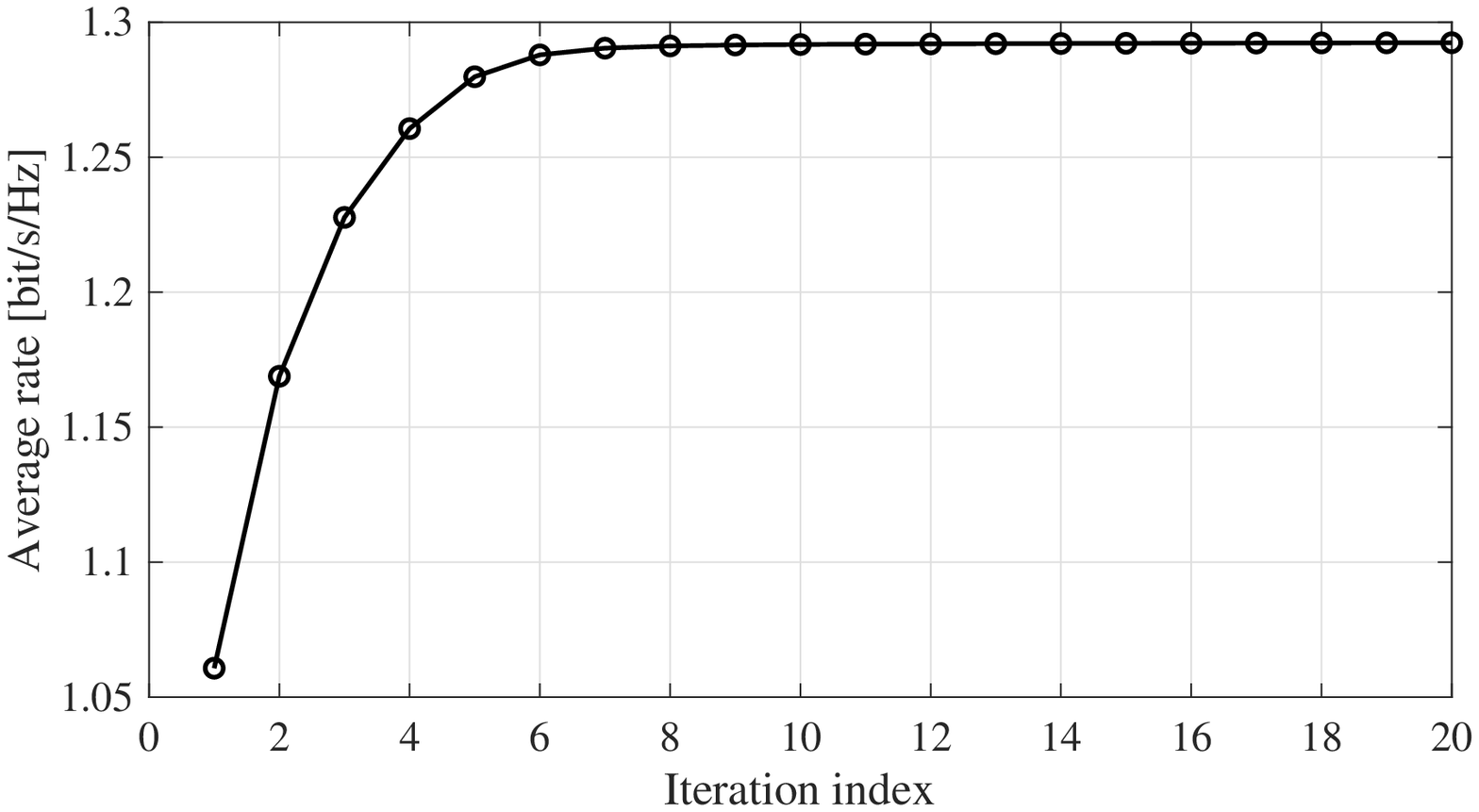}\label{fig:convergence}}

\end{center}
\caption{Case study for UL transmission of cellular-connected UAVs. (a) Average rate obtained in each realization. (b) CDFs for the rates of all the UAVs in the realizations. {(c) An example convergence curve of Algorithm\,1 applied to one network realization.} }
\end{figure}

{\black In the simulation, UAVs are scheduled in the same TTI (transmission time interval) and {with the same frequency resource} for all cells/sectors. A random distribution of 48 UAVs in the 48 cells is {realized 100 times.}\rmold{denoted as a topology, and totally 100 topologies are realized.}} For each {realization} Algorithm\,\ref{al1} is performed without QoS constraints. As a comparison, we also exploit the standard 3GPP LTE UL open loop power control (OLPC) mechanism \cite{8287891} with $P_0 = -90.8\,\text{dBm}$ and $\alpha = 0.8$. Fig.\,\ref{fig:avepfm} illustrates the system performance (average rate) achieve in {each realization} using the Algorithm\,\ref{al1} without QoS constraints and the 3GPP OLPC, respectively, and Fig.\,\ref{fig:cdfs} illustrates the cumulative distribution functions (CDFs) of the achieved rates of all UAVs in the realizations. It can be observed that Algorithm\,\ref{al1} can significantly increase the overall system performance compared to the OLPC scheme. However, the fairness among the UAVs is worse, as it can be observed from Fig.\,\ref{fig:cdfs} that around 40\% of UAVs are sacrificed with very low transmission rates. This is because some UAVs (e.g. at the cell edges) will cause severe interference to other UAVs if they want to achieve a better SINR, and
they are muted to maximize system performance. Nevertheless, certain QoS constraints can be set in
Algorithm\,\ref{al1} to increase the fairness. It is worth noting that assuming QoS constraints for the UAVs is non-trivial as there could be no feasible power solutions. {Setting the QoS constraints sophisticatedly is out of the scope of this paper, since we herein only aim to show that Algorithm\,\ref{al1} is also applicable with QoS constraints considered.} The minimum QoSs for the UAVs in each realization are set as that obtained from the OLPC scheme. In this way, a feasible power initialization can be easily chosen as the power allocation of OLPC. The achieved system performance (average rate) of each realization and the rate CDF of all the UAVs in the realizations are also illustrated in Fig.\,\ref{fig:avepfm} and Fig.\,\ref{fig:cdfs}, respectively. It can be observed that based on the OLPC constraints, Algorithm\,\ref{al1} can further increase the performances of system and individual UAVs, Moreover,
the CDF of OLPC in the low rate region is kept (slightly shifted to the right) in the CDF of Algorithm\,\ref{al1} assuming OLPC QoS constraints. In addition, it is easy to understand that the overall system performance with QoS constraints is lower than that without QoS constraints. The system performances averaged across the 100 {realizations} are calculated as 1.33\,bit/s/Hz and 0.64\,bit/s/Hz for Algorithm\,\ref{al1} without and with QoS constraints, respectively. Compared to 0.51\,bit/s/Hz obtained using the OLPC scheme, the system gains are 312\% and 25\%, respectively.

However, compared to the required UL speed (50\,Mbps/18\,MHz, i.e., 2.8\,bit/s/Hz) to support the enhanced UAV communication in LTE \cite[Table\,I]{8581827}, the obtained capacity may be still not enough.\footnote{Considering the bandwidth efficiency, coding gap, fast fading, etc., the practically required speed should be much higher than 2.8\,bit/s/Hz.} {We thus believe that} advanced techniques, e.g. directional antennas or beamforming
\cite{9082692}, have to be further utilized. Moreover, some schedule algorithms, e.g. the proportional fair principle \cite{4022307}, can also be applied jointly to improving the fairness without too much loss of the overall performance. {In addition, partially decentralizing the algorithm to decrease the system load is also a practically important research direction.} {Nevertheless, the case study has verified the applicability of Algorithm\,\ref{al1} for large-scale networks in 5G communications and preliminary shown the potential of Algorithm\,\ref{al1} in significantly increasing the overall performance.}

\section{Conclusions} \label{sect:conclusions}

{In this paper, we provided new insights to enable the GP-based power control scheme applicable for both small-scale and especially large-scale networks in the low SINR regime. By introducing auxiliary variables, the power control problem aiming to maximize the weighted sum rate of the system was transformed to an intuitive form. Based on the transformed form, a novel condensation method with linear complexity was proposed. The performance of the proposed algorithm was verified via simulations. Its potential when applied in 5G was also preliminary illustrated via a representative use-case of cellular-connected UAV communications. Meanwhile, the case study indeed showed the feasibility of the proposed algorithm applied in large-scale networks, hence verified the purpose of this paper.} 
\rmold{
In this paper, a standard form for the power control problem aiming to maximize the weighted sum rate of the system with power and QoS constraints was  presented by introducing auxiliary variables. Based on the standard form, a novel condensation method was proposed, which enables the solution through solving a series of GPs in the low SINR regime. Moreover, the proposed condensation method can be straightforwardly scaled with linearly increasing complexity. {\black Its performance in achieving the global optimality has been verified in a small-scale network, i.e., case\,1 in this paper.} Furthermore, by applying it to the UL transmission of cellular-UAVs, results show its potential in increasing the system performance. However, there are still issues to be addressed. For example, much higher data rates are still required to enable the {\black enhanced high-throughput uplink UAV communications.} The fairness among UAVs needs to be improved. Techniques such as beamforming and schedule principles are possible solutions together with the proposed method. In addition, partially decentralizing the algorithm to decrease the system load is also practically important. Our future work will
investigate these points thoroughly. 
}



\setlength{\itemsep}{0em}
\renewcommand{\baselinestretch}{1.08}
\patchcmd{\thebibliography}
  {\settowidth}
  {\setlength{\parsep}{0pt}\setlength{\itemsep}{0pt plus 0.1pt}\settowidth}
  {}{}

  \normalem
\bibliographystyle{IEEEtran}
\bibliography{IEEEabrv,reference}

\end{document}